\documentclass[12pt]{article}
\usepackage{epsfig}
\begin{document}

\baselineskip=17pt plus 0.2pt minus 0.1pt

\makeatletter
\@addtoreset{equation}{section}
\renewcommand{\theequation}{\thesection.\arabic{equation}}
\def\a'{\alpha'}
\def\s{\sigma}
\def\t{\tau}
\def\tension{\frac{1}{4\pi\alpha'}}
\def\sumk{\sum^{\infty}_{k=1}}
\def\ust{u_*}
\def\ae{a_{\rm eff}}
\def\phieff{\phi_{\rm eff}}
\def\calM{{\cal M}}
\def\calO{{\cal O}}
\def\calV{{\cal V}}
\def\calD{{\cal D}}
\def\p{{\partial}}
\def\nn{{\nonumber}}
\newcommand{\bea}{\begin{eqnarray}}
\newcommand{\eea}{\end{eqnarray}}
\def\bC{\mbox{\boldmath $C$}}
\def\bZ{\mbox{\boldmath $Z$}}
\def\bR{\mbox{\boldmath $R$}}
\begin{titlepage}
\title{
\begin{flushright}
{\normalsize {\tt hep-th/0305054}}\\
\end{flushright}
\vspace{1cm}
{\bf  Decay Rates of Fixed Planes and\\
Closed-string Tachyons on Unstable Orbifolds
}
}
\author{
Shin {\sc Nakamura}
\thanks{{\tt nakashin@postman.riken.go.jp}}
\\[15pt]
{\it Theoretical Physics Laboratory}\\
{\it RIKEN (The Institute of Physical and Chemical Research)}\\
{\it2-1 Hirosawa, Wako, Saitama 351-0198, Japan} 
\\[10pt]
}
\date{\normalsize May, 2003}
\maketitle
\thispagestyle{empty}

\begin{abstract}
\normalsize

We consider closed-string tachyon condensation in the
twisted sectors on the
$\bC/\bZ_{2n+1} \times \bR^{7,1}$ orbifold.
We calculate the localized energy density in the fixed plane
on the orbifold at the one-loop level, and we obtain
the decay rate per unit volume of the fixed plane
to leading order.
We show that the decay rate increases monotonically as
a function of $n$.

\end{abstract}

\end{titlepage}
\section{Introduction}

It is generally believed that string theory, or its non-perturbative
framework, has the capability of describing the dynamical 
transmutation of the spacetime structure.
For example, a mechanism of dynamical reduction in 
spacetime dimensions has been proposed in the context 
of the IIB matrix model \cite{IIB},
and the dynamical transmutation of spacetime topology has 
also been investigated in the framework of string theory.
Although we do not have a full description of the process,
several possible mechanisms for the dynamical transmutation of 
spacetime topology have been proposed in the context of
closed-string tachyon condensation on non-compact orbifolds
\cite{Adams,HKMM}.
This feature of string theory is quite attractive, because
we cannot describe the transmutation of the spacetime structure
within the framework of ordinary particle field theories.
With these matters in mind, we consider in this article
closed-string tachyon condensation on unstable orbifolds.

Adams, Polchinski and Silverstein (APS) have pointed
out that the fixed-point singularity in an unstable orbifold
will disappear through the condensation of
closed-string tachyons in the twisted sectors
on the orbifold \cite{Adams}.
The closed-string tachyons in the twisted sectors are
localized in the fixed plane,\footnote{
We refer to the set of fixed points as the `fixed plane' 
in this article,
because fixed points constitute a hyper-surface in general.
}
and this fixed plane disappears through tachyon condensation.
This process resembles the decay of D-branes through open-string
tachyon condensation on them.
Thus, one way to study closed-string tachyon condensation 
in the twisted sectors is to formulate an analogy with
open-string tachyon condensation;
for example,
Harvey, Kutasov, Martinec and Moore
proposed the quantity $g_{\mbox{\scriptsize cl}}$,\footnote{
See Ref. \cite{Nam} for another proposal for 
$g_{\mbox{\scriptsize cl}}$.} 
which characterizes
the process of closed-string tachyon condensation
in the twisted sectors through such an analogy \cite{HKMM}.
It is natural to inquire how far we can use an analogy
between the tachyon condensation of open strings and
that of closed strings in the twisted sectors.
In the case of open-string tachyon condensation,
the condensation is closely related to the decay of the
D-brane;
for example, the tension of the bosonic D-brane on which 
open-string tachyons exist is equal to the difference between 
the tree-level tachyon potentials
of an unstable vacuum and a stable vacuum.
The tension of the D-brane is essentially the localized
energy density which is produced by open strings
localized on the D-brane.
In this sense, it is important to consider the energy density 
localized in the fixed plane in the present case, too.
The tree-level localized energy density has been investigated
in several works \cite{Dabholkar,Dab-Vafa,Sarkar}.

We consider the $\bC/\bZ_{2n+1} \times \bR^{7,1}$ orbifold
for simplicity in this work, and the
localized energy density in the fixed plane.
However, we do not consider the tree-level localized energy 
density;
we focus on the localized energy density
at the one-loop level.
We extract the decay rate per unit volume of the fixed plane
at leading order from the localized energy density 
at the one-loop level by using the method proposed 
in Ref. \cite{Marcus}.
The localized energy density diverges at the one-loop level
because there exist tachyons in the twisted sectors.
However, we can obtain a finite contribution after 
application of the 
appropriate regularization proposed in Ref. \cite{Marcus}.
We then calculate the decay rate of the unstable fixed
plane by using the regularized energy density
and show that the decay rate of the fixed plane on the 
$\bC/\bZ_{2n+1} \times \bR^{7,1}$ orbifold grows 
monotonically as a function of $n$.

In the next section, we consider the one-loop amplitude
of type II superstrings on the $\bC/\bZ_{2n+1} \times \bR^{7,1}$ 
orbifold.
The one-loop amplitude corresponds to
the energy induced by vacuum fluctuations on the orbifold.
The induced energy can be divided into an
unlocalized component and a localized component in the
fixed plane.
We consider the question of how to extract the localized component
from the total one-loop amplitude. 
In section
3, we consider the one-loop correction to
the energy density of the fixed plane on the orbifold,
utilizing the method introduced in Ref. \cite{Marcus}.
We also consider the decay rate of the fixed plane.
In section 
4, we explicitly evaluate the decay rate per unit volume
of the fixed plane on the orbifold
and show that this decay rate per unit volume of the fixed plane
grows monotonically as a function of $n$.
We present conclusions and some open problems in
the last section.

\section{Localized component of the partition function\\
on a $\bC/\bZ_{2n+1} \times \bR^{7,1}$ orbifold}
\label{review}

Let us start by reviewing some of the basic properties of
type II superstrings on a 
$\bC/\bZ_{2n+1} \times \bR^{7,1}$ orbifold.
We define the $\bC/\bZ_{N} \times \bR^{7,1}$ orbifold
by identifying the 8-9 plane under the rotation $R$
given by
\begin{eqnarray}
R=e^{2\pi i (1+\frac{1}{N})J_{89}},
\end{eqnarray}
where $J_{89}$ is the rotation generator.
The 8-9 plane becomes a cone with deficit angle 
$2\pi(1-\frac{1}{N})$ under this identification.
The tip of the cone, the origin of the 8-9 plane ($x^{8}=x^{9}=0$),
is the fixed point that constitutes the $(7+1)$-dimensional
fixed plane in the target space.
Spacetime fermions exist if $N$ is odd, and the ground state 
of the untwisted sector is massless 
in this case \cite{Adams}.\footnote{
The extra $2\pi$ rotation in $R$ is necessary for the existence of
fermions in the untwisted sector \cite{Adams}.}
We therefore take $N$ to be a positive odd integer ($N=2n+1$)
in this paper
so that there are no tachyons in the untwisted sector.

Let us consider type II superstrings on such an orbifold.
We use the RNS formalism with the following worldsheet action:
\begin{eqnarray}
I=-\frac{1}{4\pi}\int d^{2}\sigma
\left(
\frac{1}{\alpha'}\partial_{a} X^{\mu} \partial^{a} X_{\mu}
-i\bar{\psi}^{\mu}\rho^{a}\partial_{a}\psi_{\mu}
\right).
\label{wsaction}
\end{eqnarray}
It is convenient to represent $X^{8}$, $X^{9}$, $\psi^{8}$ and
$\psi^{9}$ as complex fields according to the following:
\begin{eqnarray}
X&=&\frac{X^{8}+iX^{9}}{\sqrt{2}},\ \ \ 
\bar{X}=\frac{X^{8}-iX^{9}}{\sqrt{2}},\\
\psi&=&\frac{\psi^{8}+i\psi^{9}}{\sqrt{2}},\ \ \ 
\bar{\psi}=\frac{\psi^{8}-i\psi^{9}}{\sqrt{2}}.
\end{eqnarray}
The boundary conditions in the $m$-th twisted sector
($m=0,\cdots,N-1$)\footnote{
$m=0$ corresponds to the untwisted sector.
} are
\begin{eqnarray}
\psi(\sigma_{1}+2\pi,\sigma_{0})
&=&
e^{2\pi i(1+\frac{1}{N})m} \psi(\sigma_{1},\sigma_{0}),
\:\:\:\:\:\:\: (\mbox{R sector})
\nonumber \\
\psi(\sigma_{1}+2\pi,\sigma_{0})
&=&
-e^{2\pi i(1+\frac{1}{N})m} \psi(\sigma_{1},\sigma_{0}),
\:\:\:\: (\mbox{NS sector})
\nonumber \\
X(\sigma_{1}+2\pi,\sigma_{0})
&=&
e^{2\pi i(1+\frac{1}{N})m} X(\sigma_{1},\sigma_{0}) \ .     
\label{boundary}
\end{eqnarray}
The one-loop amplitude of the strings on the orbifold is
given by
\begin{eqnarray}
A_{\mbox{\scriptsize string}}&\propto&
V_{8}\int_{\cal F}\frac{d\tau_{1} d\tau_{2}}{2\tau_{2}}
(4\pi^{2}\alpha' \tau_{2})^{-4}
Z(\tau),
\label{Ztau}
\\
Z(\tau)&=&
\sum^{N-1}_{l=0}\sum^{N-1}_{m=0}
\frac{|\theta_{3}(\nu_{lm}|\tau)\theta_{3}(\tau)^{3}
-\theta_{2}(\nu_{lm}|\tau)\theta_{2}(\tau)^{3}
-\theta_{4}(\nu_{lm}|\tau)\theta_{4}(\tau)^{3}
|}
{4N|\eta(\tau)|^{18}|\theta_{1}(\nu_{lm}|\tau)|^{2}}
\nonumber \\
&=&
\sum^{N-1}_{l=0}\sum^{N-1}_{m=0}
\frac{|\theta_{1}(\frac{\nu_{lm}}{2}|\tau)|^{8}}
{N|\eta(\tau)^{9}\theta_{1}(\nu_{lm}|\tau)|^{2}}
=
\frac{1}{N}\sum^{N-1}_{l=0}\sum^{N-1}_{m=0}
Z_{l,m}(\tau),
\end{eqnarray}
where $\tau=\tau_{1}+i\tau_{2}$ is the moduli of the torus,
and $\nu_{lm}=\frac{N+1}{N}(l-m\tau)$ \cite{Dab-2,Lowe,Takayanagi}.
The quantity $V_{8}$ is the volume of the $\bR^{7,1}$ part of the
spacetime, which is equal to the volume of the fixed plane, and
${\cal F}$ is the fundamental region for $\tau$, which is
typically chosen as that satisfying 
$-\frac{1}{2}\le \tau_{1} \le \frac{1}{2}$
and $1\le |\tau|$.
We define $Z_{l,m}(\tau)$ as
\begin{eqnarray}
Z_{l,m}(\tau)=
\frac{|\theta_{1}(\frac{\nu_{lm}}{2}|\tau)|^{8}}
{|\eta(\tau)^{9}\theta_{1}(\nu_{lm}|\tau)|^{2}} \ .
\label{Zlm}
\end{eqnarray}
The meaning of $m$ is that $Z_{l,m}(\tau)$ comes from the
$m$-th twisted sector.
The sum over $l$ should be regarded as the $\bZ_{N}$ projection
$\frac{1}{N}\sum^{N-1}_{l=0}R^{l}$, which extracts 
the invariant states under the action of $R$.

Some comments regarding the untwisted part of the
partition function are in order. The quantity
$\frac{1}{N}\sum^{N-1}_{l=0}Z_{l,0}(\tau)$ is the
contribution from the untwisted sector,
although it should be noted that there is a slight difference 
between $Z_{0,0}(\tau)$ and $Z_{l\neq 0,0}(\tau)$.
Let us first consider $Z_{0,0}(\tau)$.
The denominator in Eq. (\ref{Zlm}) includes 
$|\theta_{1}(0|\tau)|^{2}=0$, which might cause a divergence.
This divergence is due to the zero modes of the untwisted
sector and represents the infinite volume of the
8-9 plane, which is perpendicular to the 
fixed plane \cite{Lowe,Takayanagi}.\footnote{
In the present model, however, 
 $Z_{0,0}(\tau)$ 
 eventually vanishes, due to the effect of numerator 
 $|\theta_{1}(0|\tau)|^{8}=0$, which
 represents spacetime supersymmetry in the original type II
 superstring models.}
Therefore, we have a ten-dimensional volume factor in the
$l=m=0$ part of $A_{\mbox{\scriptsize string}}$.
This can be easily understood, because the orbifold with
$N=1$, where $Z(\tau)=Z_{0,0}(\tau)$, 
is not actually an orbifold but an ordinary
flat spacetime in which all contributions of strings
are extended into the ten-dimensional spacetime. 
By contrast, there is no such divergent component
in $Z_{l\neq 0,0}(\tau)$, and the volume factor is just $V_{8}$
in that case. 
This implies that $Z_{l\neq 0,0}(\tau)$ represents a part of 
the quantity localized in the $(7+1)$-dimensional subspace
of the target space.
We find that this $(7+1)$-dimensional subspace is the
fixed plane through the following consideration.

The one-loop amplitude corresponds to the
vacuum energy induced by vacuum fluctuations
at the one-loop level.
Let us consider the vacuum energy induced by the twisted-sector
closed strings at the one-loop level.
Strings in the twisted sectors are localized in the fixed plane,
because their center-of-mass coordinates are located in
the fixed plane. 
Therefore, the vacuum energy induced by the twisted-sector
closed strings is localized in the fixed plane.
It may be thought that the contribution of the twisted sectors
just gives the localized
component of the induced vacuum energy.
However, the contribution of the twisted sectors alone
is not enough to construct the localized vacuum energy.
This is because
the contribution of the twisted sectors to $Z(\tau)$
is given by
$\frac{1}{N}\sum^{N-1}_{l=0}\sum^{N-1}_{m=1}Z_{l,m}(\tau)
\equiv Z^{\mbox{\scriptsize tw}}(\tau)$,
although 
$\frac{d\tau_{1}d\tau_{2}}{\tau_{2}^{5}}
Z^{\mbox{\scriptsize tw}}(\tau)$ 
itself is not modular invariant.
The only way to make it modular invariant is to incorporate
$\frac{1}{N}Z_{l\neq 0,0}(\tau)$ into
$Z^{\mbox{\scriptsize tw}}(\tau)$
and to consider
\begin{eqnarray}
Z^{\mbox{\scriptsize local}}(\tau)
&\equiv&
\frac{1}{N}
\sum^{N-1}_{l=0}\sum^{N-1}_{m=1}Z_{l,m}(\tau)
+\frac{1}{N} Z_{l\neq 0,0}(\tau)
\nonumber \\
&=&
\frac{1}{N}
\sum_{\{l,m\}\neq \{0,0\}}Z_{l,m}(\tau)
.
\end{eqnarray}
It is thereby found that
$\frac{d\tau_{1}d\tau_{2}}{\tau_{2}^{5}}Z^{\mbox{\scriptsize local}}(\tau)$
is modular invariant.
Of course, the localized vacuum energy is a physical quantity
and should be represented in a modular invariant way.

From the above considerations, it is found that 
the localized component of $A_{\mbox{\scriptsize string}}$
in the fixed plane can be represented as
\begin{eqnarray}
A_{\mbox{\scriptsize string}}^{\mbox{\scriptsize local}}
\propto
V_{8}\int_{\cal F}\frac{d\tau_{1} d\tau_{2}}{2\tau_{2}}
(4\pi^{2}\alpha' \tau_{2})^{-4}
Z^{\mbox{\scriptsize local}}(\tau).
\end{eqnarray}
Therefore, $\frac{1}{N}Z_{l\neq 0,0}(\tau)$ should also be regarded 
as a part of the localized component of the partition function
in the fixed plane.

Let us consider the meaning of the word ``localized" we have
used here.
It is known that closed strings in the twisted sectors
can sweep the spacetime far from the fixed plane,
although their center-of-mass coordinates are always in the
fixed plane.
However, twisted strings have to be stretched in order to
reach points far from the fixed plane, and this costs
energy.
Therefore, the main contribution of the twisted strings
comes from the diagrams that sweep the vicinity of the
fixed plane.
This is the meaning of the ``localized" contribution.

\begin{figure}
\centerline{\includegraphics{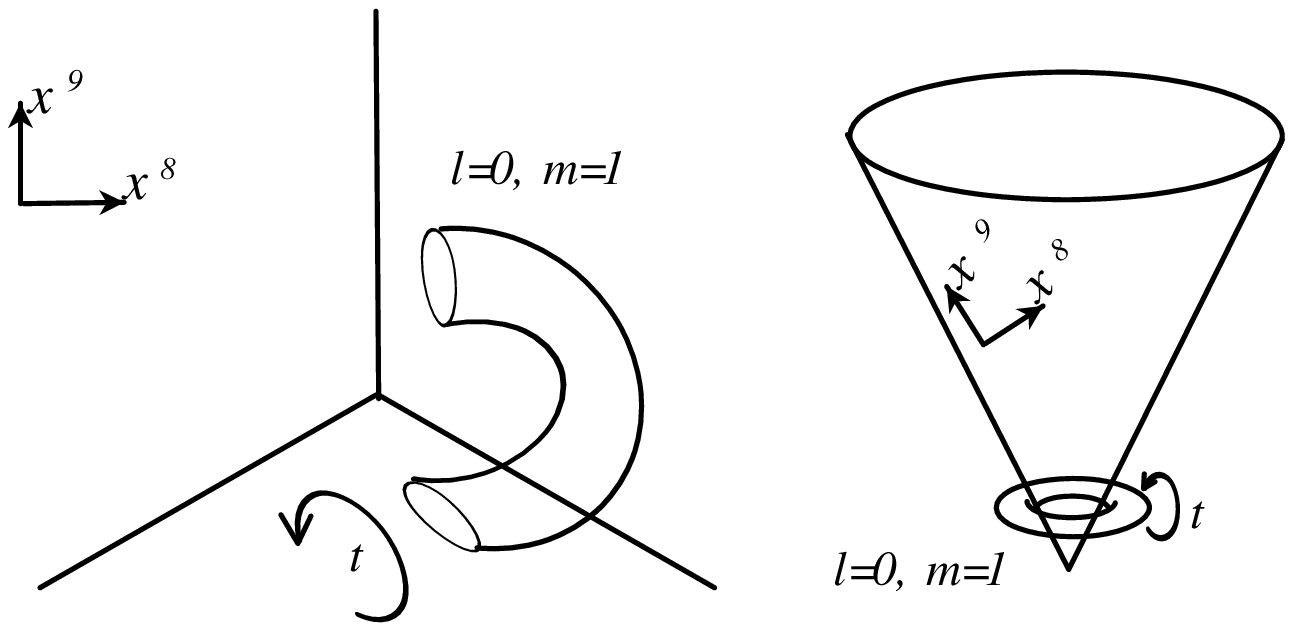}}
  \caption{All one-loop diagrams of strings in the twisted sectors
  are twisted around the fixed plane, which is located at the tip
  of the cone. 
  (The diagram corresponding to $Z_{l=0,m=1}$ on the 
  $\bC/\bZ_{3}\times \bR^{7,1}$ orbifold 
  is shown in this figure, where $t$ denotes the proper time.)}
  \label{fig:twist.eps}
%
\centerline{\includegraphics{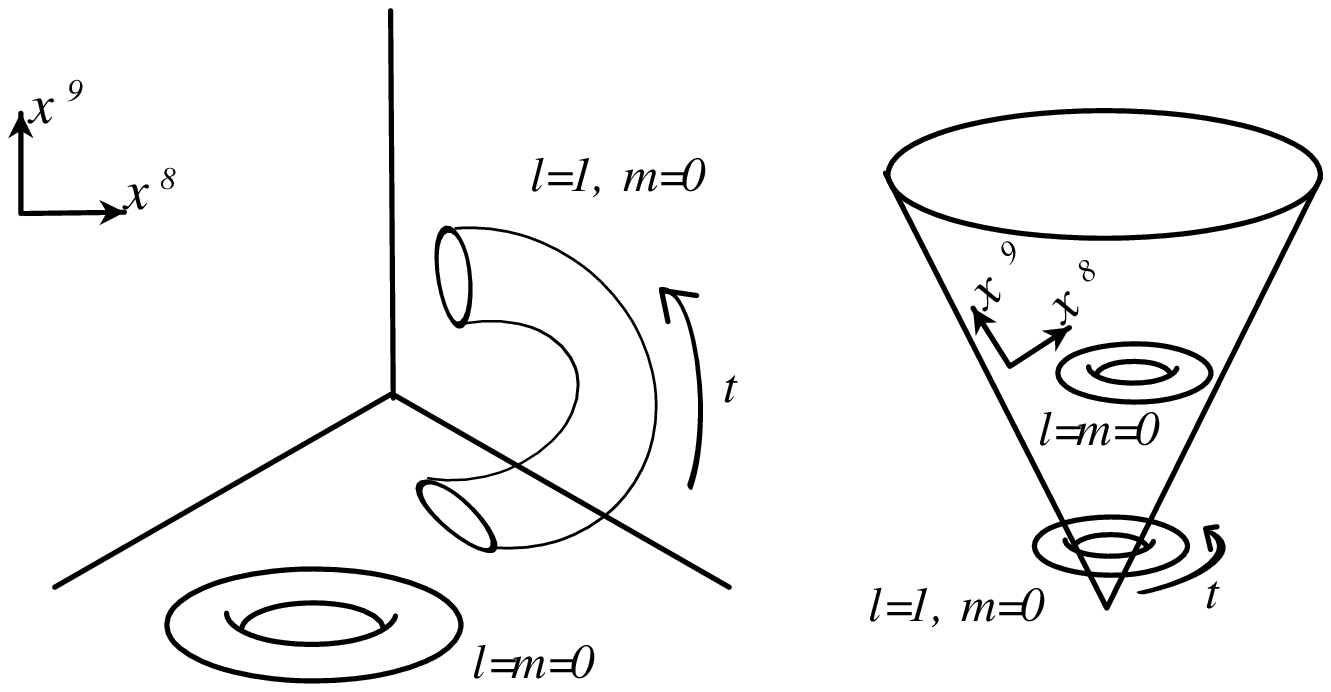}}
  \caption{Some of the one-loop diagrams of untwisted strings
  are also twisted around the fixed plane. 
  (Diagrams corresponding to $Z_{l=0,m=0}$ and $Z_{l=1,m=0}$
   on the $\bC/\bZ_{3}\times \bR^{7,1}$ orbifold 
  are shown in this figure. The $l\neq 0$ case is twisted.
  $t$ denotes the proper time.)}
  \label{fig:untwist.eps}
\end{figure}

In the case of untwisted strings which contribute to
$Z_{l\neq 0,0}(\tau)$, their center-of-mass coordinates
are not restricted to the fixed plane. 
However, their one-loop diagrams are twisted around
the fixed plane.
The contribution of the one-loop diagrams that 
sweep points far from the fixed plane is therefore smaller
than the contribution of those that sweep the vicinity 
of the fixed plane.
Thus, the main contribution of $Z_{l\neq 0,0}(\tau)$
comes from the diagrams that sweep the vicinity of 
the fixed plane.\footnote{
The author thanks S. Sugimoto for discussions
on this topic.}
With this realization, we find that it is natural to
regard $Z^{\mbox{\scriptsize local}}(\tau)$ as the contribution
to the induced vacuum energy localized in the fixed plane.
The situation described here is schematically depicted in
Figs. \ref{fig:twist.eps} and \ref{fig:untwist.eps}.

From the above discussion, we find that
the vacuum energy density localized in the
fixed plane at the one-loop level is given by
\begin{eqnarray}
\rho^{\mbox{\scriptsize local}}
=-\frac{1}{2}
\int_{\cal F}\frac{d\tau_{1} d\tau_{2}}{\tau_{2}}
(4\pi^{2}\alpha' \tau_{2})^{-4}
Z^{\mbox{\scriptsize local}}(\tau).
\label{rho-local}
\end{eqnarray}
We have fixed the overall normalization of the right-hand side
of Eq. (\ref{rho-local}) so that the normalization is the same
as that of the corresponding particle field theory
in the low-energy limit.

It is easily found that $Z_{0,0}(\tau)=0$ in the present model.
Thus, $Z^{\mbox{\scriptsize local}}(\tau)$ is equal to
$Z(\tau)$ in our model.
This implies that the induced vacuum energy at the one-loop
level in our model arises only in the vicinity of the fixed plane.

\section{One-loop localized energy and 
decay rate of the fixed plane}

APS have pointed out that the 
$\bC/\bZ_{2n+1} \times \bR^{7,1}$ 
orbifold is dynamically transmuted into the
$\bC/\bZ_{2n'+1} \times \bR^{7,1}$ orbifold,
where $n>n'$, through closed-string tachyon condensation
in the twisted sectors on the original orbifold.
The final result of this process is ordinary flat spacetime 
without orbifolding ($n'=0$), 
which possesses supersymmetry and no tachyons.
In other words, the fixed plane on the initial orbifold
disappears through the condensation of the twisted-sector 
closed-string tachyons localized in the fixed plane.
This is reminiscent of the decay process of bosonic D-branes, 
which is brought about by open-string tachyon condensation.
It is therefore natural to formulate an analogy between a
D-brane and the fixed plane.

In the sense described above, 
consideration of the ``tension" of the fixed plane 
on the orbifold is of interest.
The tension of the D-brane is the energy density localized in 
the D-brane.
Therefore, it is quite important to consider the energy density 
localized in the fixed plane.
Several proposals to define such an energy density at the
classical level have been made in Refs. \cite{Dabholkar,Dab-Vafa,Sarkar}.

We have seen that $\rho^{\mbox{\scriptsize local}}$ 
can also be regarded as the energy density localized in 
the fixed plane induced by vacuum fluctuations.
However, this is a quantity at the one-loop level of string
theory and should be regarded as the one-loop correction
to the classical energy density.
Although the main part of the energy density of the fixed plane
is given by the tree-level energy density, we consider the
one-loop energy density $\rho^{\mbox{\scriptsize local}}$ 
in this study.

We should point out that
$\rho^{\mbox{\scriptsize local}}$ defined by Eq. (\ref{rho-local})
has an IR divergence due to 
the twisted-sector tachyons, and we need an appropriate
regularization in order to obtain a finite result.
Fortunately, an appropriate regularization method with
analytic continuation has been proposed in Ref. \cite{Marcus}.

\subsection{Analytic continuation of $\rho^{\mbox{\scriptsize local}}$} 
\label{Analytic}

The presence of an IR divergence does not necessarily imply the
failure of the theory.
The IR divergence in $\rho^{\mbox{\scriptsize local}}$
due to tachyons only indicates the instability 
of the vacuum, and we can extract physically sensible
quantities from it.
A method to regularize the IR divergence due to tachyons
in the one-loop amplitude has been proposed in Ref. 
\cite{Marcus}\footnote{
See also Ref. \cite{S.Weinberg}.
}.\footnote{
This method has also been applied to the regularization of
the IR divergence due to
open-string tachyons in the annulus diagram
\cite{Bardakci,Craps,Nakamura}.}

The IR divergence in $\rho^{\mbox{\scriptsize local}}$ 
has the form
\begin{eqnarray}
\sim \sum_{i}\int^{\infty}_{\Lambda} \frac{d\tau_{2}}{\tau_{2}^{5}}
e^{-\pi \alpha' \tau_{2} M_{i}^{2}},
\end{eqnarray}
where $M_{i}^{2}<0$ is the mass squared of the tachyon
labeled $i$ that propagates in the one-loop diagram.
Here, $\Lambda$ denotes the UV cutoff.
This divergence can be regularized as follows 
by use of analytic continuation:
\begin{eqnarray}
\int^{\infty}_{\Lambda} \frac{d\tau_{2}}{\tau_{2}^{5}}
e^{-\pi \alpha' \tau_{2} (M_{i}^{2}-i\epsilon)}
\equiv
(\pi\alpha' M_{i}^{2})^{4}
\int^{\infty}_{M_{i}^{2}\Lambda-i\epsilon}
\frac{d\tau_{2}}{\tau_{2}^{5}}
e^{-\tau_{2}}. 
\:\:\:\:\:\:\:\: \mbox{(for $M_{i}^{2}<0$)}
\label{regul-part}
\end{eqnarray}
However, the regularized amplitude becomes complex in general.

\subsubsection{Imaginary part}
The imaginary part of Eq. (\ref{regul-part}) can be obtained as
\begin{eqnarray}
{\rm Im}
\left\{
\int^{\infty}_{\Lambda} \frac{d\tau_{2}}{\tau_{2}^{5}}
e^{-\pi \alpha' \tau_{2} (M_{i}^{2}-i\epsilon)}
\right\}
=\frac{\pi(-\pi\alpha'M_{i}^{2})^{4}}{\Gamma(1+4)}.
\label{imaginary-part}
\end{eqnarray}
Thus, the imaginary part of $\rho^{\mbox{\scriptsize local}}$
after regularization is
\begin{eqnarray}
{\rm Im}\{
\rho^{\mbox{\scriptsize local}}
\}
=-\frac{1}{2}\frac{1}{4^{4}4! \pi^{3}}
\sum_{i}(-M_{i}^{2})^{4},
\end{eqnarray}
where the sum over $i$ runs for all the tachyonic states.

This imaginary part has a physical meaning
\cite{Marcus,Weinberg-Wu}: The quantity
$\rho^{\mbox{\scriptsize local}}V_{8}$ can be regarded as
the one-loop correction to the effective action
of the string theory on the orbifold.
Thus, 
\begin{eqnarray}
\Gamma_{n}&=&-2{\rm Im}\{\rho^{\mbox{\scriptsize local}}\}
\nonumber \\
&=&
\frac{1}{4^{4}4! \pi^{3}}\sum_{i}(-M_{i}^{2})^{4}
\label{Gamma-n}
\end{eqnarray}
should correspond to the decay rate of the vacuum
of the $\bC/\bZ_{2n+1} \times \bR^{7,1}$ orbifold.
The subscript $n$ of $\Gamma_{n}$ indicates that the discrete
group we have used in the construction of this orbifold
is $\bZ_{2n+1}$.

The precise meaning of $\Gamma_{n}$ can be understood by employing
the argument given in Ref. \cite{Weinberg-Wu} in the framework of
particle field theory. 
The decay rate per unit volume of an unstable vacuum is
defined in Ref. \cite{Weinberg-Wu} by considering 
the time evolution of the initial
vacuum state; 
it is extracted from the time dependence of the overlap 
of the initial unstable state localized at the top of the
tachyon potential with its time-developed state.
The correspondence between $\Gamma_{n}$ obtained in the 
framework of string theory and the decay rate
given in Ref. \cite{Weinberg-Wu} in particle field theory
can be understood as follows. 
Let us consider as an example a $d$-dimensional
bosonic scalar field theory. Then, the
one-loop amplitude for the bosonic particles of mass $M$ is
given by
\begin{eqnarray}
A_{\mbox{\scriptsize field}}
=-\frac{1}{2}V_{d}
\int\frac{d^{d}p}{(2\pi)^{d}}
\ln(p^{2}+M^{2}-i\epsilon).
\end{eqnarray}
If these bosonic particles are tachyons and
$M^{2}<0$, then $\ln(p^{2}+M^{2}-i\epsilon)$ has an imaginary part,
$-i\pi$, when $p^{2}<-M^{2}$.
The imaginary part of the one-loop amplitude can be easily
evaluated as
\begin{eqnarray}
{\rm Im}\{
A_{\mbox{\scriptsize field}}
\}
&=&
\frac{1}{2}V_{d}
\int\frac{d^{d}p}{(2\pi)^{d}}
\theta(-M^{2}-p^{2})\pi
\nonumber \\
&=&
\frac{1}{2}V_{d}
\frac{\pi}{(4\pi)^{d/2}}
\frac{(-M^{2})^{d/2}}{\Gamma(1+d/2)}.
\end{eqnarray}
Here, $A_{\mbox{\scriptsize field}}$ is the one-loop
correction to the effective action for the scalar field,
and
\begin{eqnarray}
\Gamma_{\mbox{\scriptsize field}}\equiv
\frac{2\:{\rm Im}\{A_{\mbox{\scriptsize field}}\}}{V_{d}}
=\frac{\pi}{(4\pi)^{d/2}}
\frac{(-M^{2})^{d/2}}{\Gamma(1+d/2)}
\label{decay-field}
\end{eqnarray}
gives the decay rate per unit volume of the unstable vacuum
given in Ref. \cite{Weinberg-Wu}\footnote{
See also Ref. \cite{Callan}.}.
Note that if we have more than one scalar field in 8-dimensional
spacetime, the generalized expression of Eq. (\ref{decay-field})
coincides with Eq. (\ref{Gamma-n}).

In the present model, there are no tachyons in the bulk,
and the instability of the vacuum is localized only 
in the fixed plane.
Furthermore,
the canonical dimension of $\Gamma_{n}$ indicates that $\Gamma_{n}$
is the decay rate per unit volume of the 7-dimensional subspace.
Therefore, it is natural to regard $\Gamma_{n}$
as the decay rate per unit volume of the fixed plane
at the one-loop level of the perturbative string theory.
The sum $\sum_{i}(-M_{i}^{2})^{4}$ depends on discrete group
$\bZ_{2n+1}$.
We investigate the $n$ dependence of $\Gamma_{n}$ in the
next section.

\subsubsection{Real part}
The real part of 
$\rho^{\mbox{\scriptsize local}}$
after the analytic continuation should be regarded as
the one-loop correction to the localized energy density
in the fixed plane.
It has been pointed out in the framework of particle field theory
that the real part of the vacuum energy after the
analytic continuation depends on the choice of the
wavepacket in which each Fourier mode of the field
is placed \cite{Weinberg-Wu}.
It is also known that if we choose the minimum uncertainty 
wavepacket, the real part of the vacuum energy 
coincides with the real part of the regularized effective action.
We assume here that we are considering the case in which
we have chosen the minimum uncertainty wavepacket
in the corresponding field theory, and we regard
the real part of $\rho^{\mbox{\scriptsize local}}$
as the one-loop correction to the vacuum energy density.

Let us consider the real part of the regularized energy density
of the fixed plane on the $\bC/\bZ_{3} \times \bR^{7,1}$ 
orbifold as an example.
We must carry out a numerical calculation to obtain the
real part of $\rho^{\mbox{\scriptsize local}}$.
The tachyonic states in this orbifold are only the ground states 
in the twisted sectors.
$\rho^{\mbox{\scriptsize local}}$ can be approximately written as
\begin{eqnarray}
\rho^{\mbox{\scriptsize local}}
\sim
-\frac{1}{2}\frac{1}{(4\pi^{2}\alpha')^{4}}
\int_{1}^{\infty} \frac{d\tau_{2}}{d\tau_{2}^{5}}
\{
6e^{2\pi\tau_{2}\cdot\frac{2}{3}}+108
+ \cdots  
\}
,
\end{eqnarray}
where the contribution of the massive modes has been omitted.
We have also ignored the integral in the region satisfying
$\tau_{2}<1$ in ${\cal F}$.
The first term in the integrand is the contribution of the tachyonic
modes, and it should be defined by using analytic continuation.
After this analytic continuation, we obtain
\begin{eqnarray}
{\rm Re}\{\rho^{\mbox{\scriptsize local}}\}
\sim -\frac{O(10^{-6})}{(\alpha')^{4}} .
\end{eqnarray}
Therefore, the one-loop contribution to the localized energy density
in the fixed plane on the $\bC/\bZ_{3} \times \bR^{7,1}$
orbifold is negative.
This suggests a positive energy density
at the tree level whose magnitude is sufficiently large to
overwhelm this negative one-loop contribution.

\section{Decay rate per unit volume of the fixed plane on 
the $\bC/\bZ_{2n+1} \times \bR^{7,1}$ orbifold}
\label{decay-rate}

In this section, we consider the $n$ dependence of $\Gamma_{n}$ 
for the $\bC/\bZ_{2n+1} \times \bR^{7,1}$ orbifold.
We need to list all the physical tachyonic 
states in the twisted sectors, because $\Gamma_{n}$ is essentially 
given by $\sum_{i}(-M_{i}^{2})^{4}$, where the sum over $i$
runs over all the tachyonic states.

The mass squared of the ground state in the $m$-th 
twisted sector is given as follows:
\begin{eqnarray}
M^{2}&=&
-\frac{2}{\alpha'}\left(1-\frac{m}{2n+1}\right),
\:\:\:\:\:\:\:\:\mbox{(for odd $m$)}  \\
M^{2}&=&
-\frac{2}{\alpha'}\frac{m}{2n+1}.
\:\:\:\:\:\:\:\:\:\:\:\:\:\:\:\:\:\:\:\:\:\:\mbox{(for even $m$)}   
\end{eqnarray}
Thus, every twisted sector has tachyons.
Furthermore, some excited states can also be tachyonic.
We find that the mass squared of such a $k$-th excited 
tachyonic
state in the $m$-th twisted sector can be written as
\begin{eqnarray}
M^{2}&=&
-\frac{2}{\alpha'}\left(1-\frac{(2k+1)m}{2n+1}\right),
\:\:\:\:\:\:\:\:\mbox{(for odd $m$)}  
\label{kth-mass-odd}\\
M^{2}&=&
-\frac{2}{\alpha'}\left(\frac{(2k+1)m}{2n+1}-2k\right),
\:\:\:\:\:\:\:\:\mbox{(for even $m$)}
\label{kth-mass-even}
\end{eqnarray}
where $k$ should be a non-negative integer satisfying
$M^{2}<0$.\footnote{
See Appendix A for details.}
An important property of Eqs. (\ref{kth-mass-odd}) 
and (\ref{kth-mass-even})
is that the tachyonic spectrum for odd $m$ is exactly the same as
the tachyonic spectrum in the $l$-th twisted sector,
where $l$ is an even integer given by $l=2n+1-m$.
Therefore, we only need to consider the tachyonic spectrum
in the $m$-th twisted sector with odd $m$, that is, we have
\begin{eqnarray}
\sum_{i}(-M_{i}^{2})^{4}
=2\sum_{j}(-M_{j}^{2})^{4},
\end{eqnarray}
where $j$ labels only the tachyonic states in the $m$-th 
twisted sector with odd $m$.
This makes the subsequent analysis simpler.

Let us consider the tachyonic spectrum in the $m$-th twisted sector 
with odd $m$, which is given by Eq. (\ref{kth-mass-odd}).
We note the following facts:
\begin{itemize}
  \item Each tachyonic state becomes more tachyonic as $n$ increases.
  \item The number of tachyonic states increases as $n$ increases.
This is because $k$ for odd $m$ is bounded as
\begin{eqnarray}
1\le 2k+1<\frac{2n+1}{m}.
\end{eqnarray}
\end{itemize}
Of course, the number $2n$ of the twisted sectors also increases
with $n$.
We can therefore conclude that $\Gamma_{n}<\Gamma_{n'}$
if $n<n'$. 
Thus, the decay rate per unit volume of the fixed plane
grows monotonically as a function of $n$. 
Note that $\Gamma_{n}$ can be written explicitly as
\begin{eqnarray}
\Gamma_{n}=
\frac{2}{4!\pi^{3}\alpha'^{4}}
\sum_{r=0}^{n-1}\sum_{k=0}^{[\frac{n-r}{2r+1}]}
\left(1-\frac{(2k+1)(2r+1)}{(2n+1)}\right)^{4},
\label{explicit-Gn}
\end{eqnarray}
where we have defined $r$ as $m=2r+1$.
Here, $[\frac{n-r}{2r+1}]$ denotes the integer part of
$\frac{n-r}{2r+1}$.

Let us now consider the behaviour of $\Gamma_{n}$ in
the large $n$ limit.
We find that
\begin{eqnarray}
\Gamma_{n}
\sim
\frac{2}{4!\pi^{3}\alpha'^{4}}
\frac{1}{20}(2n+1)\ln (2n+1),
\:\:\:\:\:\:\:\:(n\to \infty)
\label{largeN-main}
\end{eqnarray}
and the growth of $\Gamma_{n}$
is not bounded.
The derivation of Eq. (\ref{largeN-main})
is given in Appendix B.

\section{Conclusions and discussion}

We have considered the energy density localized in the fixed plane
on a $\bC/\bZ_{2n+1} \times \bR^{7,1}$ orbifold at the one-loop level
and have seen that the non-vanishing component of the one-loop
amplitude on this orbifold gives the 
localized vacuum energy in the fixed plane at the one-loop level.
This implies that a part of the contribution from the untwisted 
sector, as well as the contribution from the twisted sectors,
should be regarded as the localized contribution in the
fixed plane. 

We used analytic continuation to regularize the IR
divergence in the one-loop energy density and
calculated the leading-order decay rate per unit volume 
of the fixed plane by using the imaginary part of the regularized 
energy density.
We found that the decay rate per unit volume of 
the fixed plane on the $\bC/\bZ_{2n+1} \times \bR^{7,1}$ orbifold
decreases monotonically as a function of $n$, and finally it
reaches zero at $n=0$ (note that $\Gamma_{0}=0$).

According to the observation of APS, 
the $\bC/\bZ_{2n+1} \times \bR^{7,1}$
orbifold decays into a flat spacetime through closed-string
tachyon condensation in the twisted sectors.
One of the decay paths that was investigated by APS is that in which
$n$ decreases monotonically through tachyon condensation until 
it finally reaches zero, which corresponds to a flat spacetime.
The results of this work indicate that
$\Gamma_{n}$ decreases monotonically through this process
and finally becomes zero.
This property of $\Gamma_{n}$ tempts us to find a connection of
$\Gamma_{n}$ to the total transition rate from the
$\bC/\bZ_{2n+1} \times \bR^{7,1}$ orbifold to a flat spacetime
through tachyon condensation.
Unfortunately, there does not currently exist a theoretical framework
which rigorously describes the transition of a spacetime structure.
$\Gamma_{n}$ is defined in each perturbative string theory
on each orbifold.
Therefore we cannot immediately conclude that $\Gamma_{n}$ is
the transition rate from the $\bC/\bZ_{2n+1} \times \bR^{7,1}$ 
orbifold to a flat spacetime.
The relationship between $\Gamma_{n}$ presented here and
the transition rate of the spacetime structure should be
studied in future works.

The method we have utilized to calculate the regularized
$\rho^{\mbox{\scriptsize local}}$ is also applicable
to other orbifolds that have been considered in the
context of closed-string tachyon condensation in the 
twisted sectors.
It would be interesting to examine whether or not the decay rates of
the fixed planes always decrease monotonically along the 
decay paths of the orbifolds.
It would also be interesting to consider whether there is some
connection between each tachyonic mode and each decay branch
from the viewpoint of the decay rate considered in the present work.

The relationship between the transition rate from the initial
orbifold to the flat spacetime and the decay rate we have calculated
in this paper is not clear at this stage.
However, further consideration of the results of the present work may
yield some useful information concerning the construction
of a formulation that describes the transmutation
of the spacetime structure.
\vspace{1cm}

\noindent
{\large\bf Acknowledgments}

The author thanks members of the Theoretical
Physics Laboratory at RIKEN, especially M. Hayakawa, H. Kawai, 
Y. Shibusa, M. Tachibana and T. Tada for fruitful discussions 
and comments.
He also thanks J. Ambj\o rn, P. Olesen, T. Shimada
and S. Sugimoto for valuable discussions and comments during
his stay at The Niels Bohr Institute.
The author also thanks the participants at the
Sapporo Winter School 2003 and
members of the Institute of Theoretical Physics of 
Rikkyo University for their comments and discussions.
He also wishes to thank M. Kato and H. Hata for 
important comments on the present work
during the YITP workshop YITP-W-03-07, ``Quantum Field Theory 2003'', 
held at the Yukawa Institute for Theoretical Physics 
at Kyoto University.

\newpage
\appendix

\noindent
{\bf\Large Appendix}

\section{Tachyonic spectrum}

In this appendix, we show that the tachyonic spectrum in Eqs.
(\ref{kth-mass-odd}) and (\ref{kth-mass-even}) can 
be extracted from the
partition function on the orbifold.
To begin, let us rewrite $Z_{l,m}(\tau)$ as
\begin{eqnarray}
Z_{l,m}(\tau)=
\frac{|\theta_{1}(\frac{\nu_{lm}}{2}|\tau)|^{8}}
{|\eta(\tau)^{9}\theta_{1}(\nu_{lm}|\tau)|^{2}}
=
e^{-2\pi \tau_{2}(1+\frac{1}{N})m}
F_{1}F_{2}F_{3}F_{4}F_{5},
\end{eqnarray}
where
\begin{eqnarray}
F_{1}&\equiv&
\prod_{k=1}^{\infty}(1-q^{k})^{6}(1-\bar{q}^{k})^{6}
=
\prod_{k=1}^{\infty}
\left(\sum_{j=0}^{\infty}q^{kj}\right)^{6}
\left(\sum_{j=0}^{\infty}\bar{q}^{kj}\right)^{6},\\
F_{2}&\equiv&G_{1}G_{2},\\
&&G_{1}\equiv
\prod_{k=1}^{k_{*}}
\left|(1-(-1)^{l}
e^{\pi i \frac{l}{N}}q^{k-1-(1+\frac{1}{N})\frac{m}{2}})^{4}
\right|^{2},
\label{G1}\\
&&G_{2}\equiv
\prod_{k=k_{*}+1}^{\infty}
\left|(1-(-1)^{l}
e^{\pi i \frac{l}{N}}q^{k-1-(1+\frac{1}{N})\frac{m}{2}})^{4}
\right|^{2},
\label{G2}\\
F_{3}&\equiv&
\prod_{k=1}^{\infty}
\left|(1-(-1)^{l}
e^{-\pi i \frac{l}{N}}q^{k+(1+\frac{1}{N})\frac{m}{2}})^{4}
\right|^{2},\\
F_{4}&\equiv&
\prod_{k=1}^{\infty}
\left|(1-e^{2\pi i \frac{l}{N}}q^{k-1-m(1+\frac{1}{N})})^{-1}
\right|^{2},
\nonumber \\
&=&G_{3}G_{4},\\
&&G_{3}\equiv
\prod_{k=1}^{m+1}
\left|\sum_{j=0}^{\infty}
  e^{-2\pi i \frac{l}{N}(j+1)}q^{-(k-m-1-\frac{m}{N})(j+1)}
\right|^{2},\\
&&G_{4}\equiv
\prod_{k=m+2}^{\infty}
\left|\sum_{j=0}^{\infty}
  e^{2\pi i \frac{l}{N}j}q^{(k-m-1-\frac{m}{N})j}
\right|^{2},\\
F_{5}&\equiv&
\prod_{k=1}^{\infty}
\left|(1-e^{-2\pi i \frac{l}{N}}q^{k+m(1+\frac{1}{N})})^{-1}
\right|^{2}
\nonumber \\
&=&
\prod_{k=1}^{\infty}
\left|\sum_{j=0}^{\infty}
  e^{-2\pi i \frac{l}{N}j}q^{(k+m(1+\frac{1}{N}))j}
\right|^{2}
.
\end{eqnarray}
The quantity $k_{*}$ in Eqs. (\ref{G1}) and (\ref{G2}) is
$\frac{m}{2}+1$ for even $m$ and
$\frac{m+1}{2}$ for odd $m$.
We define $q$ as $q\equiv e^{2\pi i\tau}$.

$Z_{l,m}(\tau)$ can then be rewritten in the form
\begin{eqnarray}
Z_{l,m}(\tau)=\sum_{i}e^{2\pi\tau_{2}h_{i}}f(\tau_{1}),
\end{eqnarray}
where the sum over $i$ runs for all the states.
The contribution of some states to the one-loop amplitude
should vanish after integrating over $\tau_{1}$ in the
amplitude when $\tau_{2}$ is sufficiently
large; this gives the level-matching condition.
All we then have to do is to list the states that survive 
the $\tau_{1}$ integration and choose
positive $h_{i}$ that correspond to the tachyonic states.

Let us consider the most tachyonic state.
The most tachyonic contribution that satisfies
the level-matching condition can be extracted by
picking up $1$ from $F_{1}$, $F_{3}$, $F_{5}$, $G_{2}$
and $G_{4}$, by picking up 
$\prod_{k=1}^{k_{*}}(q\bar{q})^{4(k-1-(1+\frac{1}{N})\frac{m}{2})}$
from $G_{1}$, and by
picking up the terms with $j=0$ from $G_{3}$.
This yields the following as the most tachyonic term:
\begin{eqnarray}
e^{-2\pi\tau_{2}(1+\frac{1}{N})m}
e^{-4\pi\tau_{2}4\sum_{k=1}^{k_{*}}(k-1-(1+\frac{1}{N})\frac{m}{2})}
e^{-4\pi\tau_{2}(-1)\sum_{k=1}^{m+1}(k-m-1-\frac{m}{N})}
\nonumber \\
=e^{2\pi\tau_{2}h}.
\end{eqnarray}
Here, we have
\begin{eqnarray}
h=
\left\{
  \begin{array}{cl}
    \frac{m}{N}   & (\mbox{ $m$: even})   \\
    1-\frac{m}{N}   & (\mbox{ $m$: odd})   \\
  \end{array}
\right.
,
\end{eqnarray}
which represents the correct ground states.

Excited states can also be tachyonic in general.
The terms corresponding to excited tachyonic states 
can be extracted from $Z_{l,m}(\tau)$ by simply changing 
the terms picked up from $G_{3}$ and $G_{4}$.
If we change the terms picked up from one of the other parts
($F_{1}$, $F_{3}$, $F_{5}$, $G_{1}$ or $G_{2}$),
the terms corresponding to the excited states
can give either massless or massive contributions.
It can also be shown that we should pick up 
$e^{-2\pi \tau_{2} \frac{2m}{N}(j+1)}
\prod_{k=1}^{m}e^{4\pi \tau_{2}(k-m-1-\frac{m}{N})}$ 
from $G_{3}$ and
$1$ from $G_{4}$ if $m$ is odd, while
we should pick up
terms with $j=0$ from $G_{3}$ and
$e^{-2\pi \tau_{2} 2(1-\frac{m}{N})j}$ from $G_{4}$
if $m$ is even.
These considerations lead to the following possible values of 
$h_{i}$ for the tachyonic contribution:
\begin{eqnarray}
h_{i}=
\left\{
  \begin{array}{cc}
   \frac{m}{N}-2(1-\frac{m}{N})j    & (\mbox{ $m$: even})    \\
   1-\frac{m}{N}-2\frac{m}{N}j    &  (\mbox{ $m$: odd})  \\
  \end{array}
\right.
.
\label{Tspectrum}
\end{eqnarray}
Here, $j$ is a non-negative integer.
This gives the tachyonic spectrum presented in 
Eqs. (\ref{kth-mass-odd}) and (\ref{kth-mass-even}).
We can also show that the states corresponding to Eq.
(\ref{Tspectrum}) satisfy the level-matching condition.

\section{Derivation of Eq. (\ref{largeN-main})}

Here we evaluate the large $n$ behaviour of Eq. (\ref{explicit-Gn}).
We first decompose the sum over $r$ in Eq. (\ref{explicit-Gn})
into two parts,
the sum in the region $0\le r \le [\frac{n-r}{2r+1}]$
and that in the region $[\frac{n-r}{2r+1}]+1 \le r \le n-1$.
Explicitly, we write
\begin{eqnarray}
&&\sum_{r=0}^{n-1}\sum_{k=0}^{[\frac{n-r}{2r+1}]}
\left(1-\frac{(2k+1)(2r+1)}{2n+1}\right)^{4}
\nonumber \\
&=&
\sum_{r=[\frac{n-1}{3}]+1}^{n-1}
\left(1-\frac{2r+1}{2n+1}\right)^{4}
+
\sum_{r=0}^{[\frac{n-1}{3}]}
\sum_{k=0}^{[\frac{n-r}{2r+1}]}
\left(1-\frac{(2k+1)(2r+1)}{2n+1}\right)^{4},
\label{decompose}
\end{eqnarray}
where we have used the fact that
$[\frac{n-r}{2r+1}]=0$ if $r>\frac{n-1}{3}$.

Let us consider the first term on the right-hand side of
Eq. (\ref{decompose}).
If $s$ is a positive integer, we have
\begin{eqnarray}
&&\sum_{r=s}^{n-1}
\left(1-\frac{2r+1}{2n+1}\right)^{4}
\nonumber \\
&&=\frac{8(1+2n-2s)(n-s)(1+n-s)(-1+3n+3n^{2}-3s-6ns+3s^{2})}
{15(2n+1)^{4}}
\nonumber \\
&&\equiv F_{(1)}(s,n).
\label{F1S}
\end{eqnarray}
We use Eq. (\ref{F1S}) as the formal definition of $F_{(1)}(s,n)$,
even for the case in which $s$ is not an integer.
We can show that
\begin{eqnarray}
F_{(1)}\left(\frac{n-1}{3},n\right)
>F_{(1)}\left(\left[\frac{n-1}{3}\right]+1,n\right)
\ge F_{(1)}\left(\frac{n-1}{3}+1,n\right)
\label{F1rela}
\end{eqnarray}
for $n\ge 2$.
By using Eq. (\ref{F1rela})
and taking the limit $n \to \infty$, we find that
\begin{eqnarray}
\sum_{r=[\frac{n-1}{3}]+1}^{n-1}
\left(1-\frac{2r+1}{2n+1}\right)^{4}
\sim
\frac{16}{1215}N+G_{(1)}(N),
\:\:\:\:\:\:\:\:(n\to \infty)
\end{eqnarray}
where $G_{(1)}(N)$ can be a non-trivial function of $N=2n+1$,
although it is bounded as
$-\frac{8}{135}\le G_{(1)}(N)\le \frac{56}{405}$.

We next consider the second term on the right-hand side of
Eq. (\ref{decompose}).
We define $t\equiv \frac{n-r}{2r+1}-[\frac{n-r}{2r+1}]$.
Then, we can write
\begin{eqnarray}
\sum_{k=0}^{[\frac{n-r}{2r+1}]}
\left(1-\frac{(2k+1)(2r+1)}{2n+1}\right)^{4}
=\sum_{k=0}^{\frac{n-r}{2r+1}-t}
\left(1-\frac{(2k+1)(2r+1)}{2n+1}\right)^{4}.
\end{eqnarray}
Note that $1>t\ge  0$ and 
$\frac{n-r}{2r+1}-t>0$ if $0\le r \le \frac{n-1}{3}$.
We can show that the following holds:
\begin{eqnarray}
&&\sum_{k=0}^{\frac{n-r}{2r+1}-t}
\left(1-\frac{(2k+1)(2r+1)}{2n+1}\right)^{4}
\nonumber \\
&=&
\frac{8}{15(2n+1)^{4}(2r+1)}
\left\{
15n^{4}+6n^{5}-30n^{2}r(1+r)
\right.
\nonumber \\
&&\left.
-10n^{3}(-1+2r+2r^{2})
+n(-1-8r+6r^{2}+28r^{3}+14r^{4})
\right.
\nonumber \\
&&\left.
+r(1+r)(1+7r+7r^{2})
\right\}
\nonumber \\
&-&
\frac{8(2r+1)^{4}}{15(2n+1)^{4}}
t(-1+t)(-1+2t)(-1-3t+3t^{2})
\nonumber \\
&\equiv&
F_{(2)}(t,r,n).
\label{F2t}
\end{eqnarray}
We use Eq. (\ref{F2t}) as the formal definition of $F_{(2)}(t,r,n)$,
even in the case that $\frac{n-r}{2r+1}-t$ is not an integer.
We can show that
\begin{eqnarray}
F_{(2)}(t_{2},r,n)\ge  F_{(2)}(t,r,n)\ge  F_{(2)}(t_{1},r,n),
\:\:\:\:\:\:\:\:(0\le t<1)
\end{eqnarray}
where
\begin{eqnarray}
t_{1}&=&\frac{1}{2}+\sqrt{\frac{1}{4}-\frac{1}{\sqrt{30}}}\:\:\:\: ,\\
t_{2}&=&\frac{1}{2}-\sqrt{\frac{1}{4}-\frac{1}{\sqrt{30}}}\:\:\:\: .
\end{eqnarray}
Note that $t_{1}$ and $t_{2}$ depend on neither $n$ nor $r$.
Thus we have
\begin{eqnarray}
\sum_{r=0}^{[\frac{n-1}{3}]}
F_{(2)}(t_{2},r,n)
\ge 
\sum_{r=0}^{[\frac{n-1}{3}]}
\sum_{k=0}^{[\frac{n-r}{2r+1}]}
\left(1-\frac{(2k+1)(2r+1)}{2n+1}\right)^{4}
\ge
\sum_{r=0}^{[\frac{n-1}{3}]} 
F_{(2)}(t_{1},r,n).\:\:\:\:\:\:\:\:
\label{hasamiuchi}
\end{eqnarray}

In order to evaluate 
$\sum_{r=0}^{[\frac{n-1}{3}]}F_{(2)}$,
we define $u\equiv \frac{n-1}{3}-[\frac{n-1}{3}]$.
We can then show that
\begin{eqnarray}
&&\sum_{r=0}^{\frac{n-1}{3}-u}F_{(2)}(t,r,n)
\nonumber \\
&=&
\frac{1}{291600}
\Bigg\{
-27915+72900\ln4
\nonumber \\
&&-1280t(-1+t)(-1+2t)(-1-3t+3t^{2})
\nonumber \\
&&+120u(249-16t+160t^{3}-240t^{4}+96t^{5})
\nonumber \\
&&+n
\left(
-5190+29160\ln4 
-128t(-1+t)(-1+2t)(-1-3t+3t^{2})
\right)
\Bigg\}
\nonumber \\
&&+\frac{1}{20}(2n+1)
\left\{
\gamma+\Psi\left(\frac{2n+1}{6}+1-u\right)
\right\}
+O(1/n)
\nonumber \\
&\equiv& F_{(3)}(u,t,n)+O(1/n),
\label{F3u}
\end{eqnarray}
where $\gamma$ is Euler's constant, and
$\Psi(z)=\frac{d}{dz}\ln\Gamma(z)$.
We use Eq. (\ref{F3u}) as the formal definition of 
$F_{(3)}(u,t,n)$,
even for the case in which $\frac{n-1}{3}-u$ is not an integer.
We note that $F_{(3)}(0,t_{2},n)\ge F_{(3)}(u,t_{2},n)$
and  $F_{(3)}(u,t_{1},n)\ge F_{(3)}(1,t_{2},n)$
for $0\le u<1$ if $n$ is sufficiently large.
It can thus be shown by using Eq. (\ref{hasamiuchi}) that
\begin{eqnarray}
F_{(3)}(0,t_{2},n)
\ge 
\sum_{r=0}^{[\frac{n-1}{3}]}
\sum_{k=0}^{[\frac{n-r}{2r+1}]}
\left(1-\frac{(2k+1)(2r+1)}{2n+1}\right)^{4}
\ge
F_{(3)}(1,t_{1},n)
\label{hasamiuchi-2}
\end{eqnarray}
in the limit that $n \to \infty$.
By taking the large $n$ limit of
$F_{(3)}(0,t_{2},n)$ and $F_{(3)}(1,t_{1},n)$, 
we find that
\begin{eqnarray}
\sum_{r=0}^{[\frac{n-1}{3}]}
\sum_{k=0}^{[\frac{n-r}{2r+1}]}
\left(1-\frac{(2k+1)(2r+1)}{2n+1}\right)^{4}
\nonumber \\
\sim
\frac{1}{20}N\ln N +G_{(2)}(N)N+G_{(3)}(N),
\:\:\:\:\:\:\:\:(n\to \infty)
\label{2ndterm}
\end{eqnarray}
where we have used the fact that $z\Psi(z) \sim z\ln z -\frac{1}{2}$
for $z \to \infty$. The quantities
$G_{(2)}(N)$ and $G_{(3)}(N)$ can be non-trivial functions of $N$,
although they are bounded as follows:
\begin{eqnarray}
F_{(4)}(t_{2})\ge &G_{(2)}(N)& \ge F_{(4)}(t_{1}),\\
F_{(5)}(0,t_{2})+\frac{3}{20}\ge  &G_{(3)}(N)& 
\ge F_{(5)}(1,t_{1})-\frac{3}{20}.
\end{eqnarray}
Here, $F_{(4)}(u,t)$ and $F_{(5)}(t)$ are defined as
\begin{eqnarray}
F_{(4)}(t)
&\equiv&
\frac{-2595+14580\ln4 -64t(-1+t)(-1+2t)(-1-3t+3t^{2})}{291600}
\nonumber \\
&+&\frac{\gamma}{20}-\frac{1}{20}\ln 6,\\
F_{(5)}(u,t)
&\equiv&
\frac{1}{291600}
\Bigg\{
-25320+58320\ln4
\nonumber \\
&-&1216t(-1+t)(-1+2t)(-1-3t+3t^{2})
\nonumber \\
&+&120u(249-16t+160t^{3}-240t^{4}+96t^{5})
\Bigg\},
\end{eqnarray}
and $F_{(3)}(u,t,n)$ is given by
\begin{eqnarray}
F_{(3)}(u,t,n)=\frac{2n+1}{20}
\Psi\left(\frac{2n+1}{6}+1-u\right)
+(2n+1)F_{(4)}(t)+F_{(5)}(u,t).
\end{eqnarray}
Therefore Eqs. (\ref{F1rela}) and (\ref{2ndterm}) lead to
\begin{eqnarray}
& &\sum_{r=0}^{n-1}\sum_{k=0}^{[\frac{n-r}{2r+1}]}
\left(1-\frac{(2k+1)(2r+1)}{2n+1}\right)^{4}
\nonumber \\
& &\sim
\frac{1}{20}N\ln N +G_{(4)}(N)N+G_{(5)}(N),
\:\:\:\:\:\:\:\:(n\to \infty)
\end{eqnarray}
where $G_{(4)}(N)$ and $G_{(5)}(N)$ can be non-trivial functions 
of $N$ that are bounded as
\begin{eqnarray}
F_{(4)}(t_{2})+\frac{56}{405}
\simeq 0.14
\ge &G_{(4)}(N)& 
\ge F_{(4)}(t_{1})-\frac{8}{135}
\simeq -0.060,\\
F_{(5)}(0,t_{2})+\frac{3}{20}+\frac{16}{1215}
\simeq 0.35
\ge &G_{(5)}(N)& 
\ge F_{(5)}(1,t_{1})-\frac{3}{20}+\frac{16}{1215}
\simeq 0.16. \:\:\:\:\:\:\:\:
\end{eqnarray}
The above calculation leads to the following:
\begin{eqnarray}
\Gamma_{n}
\sim
\frac{2}{4!\pi^{3}\alpha'^{4}}
\left\{
\frac{1}{20}N\ln N+G_{(4)}(N)N+G_{(5)}(N)
\right\},
\:\:\:\:\:\:\:\:(n\to \infty)
\label{largeN}
\end{eqnarray}
where
\begin{eqnarray}
1.4 \times 10^{-1} \ge  &G_{(4)}(N)& \ge -6.0 \times 10^{-2},
\label{bound1}\\
3.5 \times 10^{-1} \ge &G_{(5)}(N)& \ge 1.6 \times 10^{-1}.
\label{bound2}
\end{eqnarray}
Therefore, the leading-order behaviour of $\Gamma_{n}$ is
given by 
$\Gamma_{n}\propto (2n+1)\ln (2n+1)$ in the large $n$ region,
and we can extract Eq. (\ref{largeN-main}) from Eq. (\ref{largeN}).


\begin{thebibliography}{99}

\newcommand{\J}[4]{{\sl #1} {\bf #2} (#3) #4}
\newcommand{\andJ}[3]{{\bf #1} (#2) #3}
\newcommand{\AP}{Ann.\ Phys.\ (N.Y.)}
\newcommand{\MPL}{Mod.\ Phys.\ Lett.}
\newcommand{\NP}{Nucl.\ Phys.}
\newcommand{\PL}{Phys.\ Lett.}
\newcommand{\PR}{Phys.\ Rev.}
\newcommand{\PRL}{Phys.\ Rev.\ Lett.}
\newcommand{\ATMP}{Adv.\ Theor.\ Math.\ Phys.}
\newcommand{\PTP}{Prog.\ Theor.\ Phys.}

\bibitem{IIB}
H. Aoki, S. Iso, H. Kawai, Y. Kitazawa and T. Tada,
``Space-Time Structures from IIB Matrix Model'',
\J{\PTP}{99}{1998}{713},
hep-th/9802085.

\bibitem{Adams}
A. Adams, J. Polchinski and E. Silverstein,
``Don't Panic! Closed String Tachyons in ALE Spacetimes'',
\J{JHEP}{0110}{2001}{029},
hep-th/0108075.

\bibitem{HKMM}
J. A. Harvey, D. Kutasov, E. J. Martinec and G. Moore,
``Localized Tachyons and RG Flows'',
hep-th/0111154.



\bibitem{Nam}
S. Nam and S. Sin,
``Condensation of Localized Tachyons and Spacetime Supersymmetry'',
hep-th/0201132.


\bibitem{Dabholkar}
A. Dabholkar,
``Tachyon Condensation and Black Hole Entropy'',
\J{\PRL}{88}{2002}{091301},
hep-th/0111004.


\bibitem{Dab-Vafa}
A. Dabholkar and C. Vafa,
``tt* Geometry and Closed String Tachyon Potential'',
\J{JHEP}{0202}{2002}{008},
hep-th/0111155.

\bibitem{Sarkar}
S. Sarkar and B. Sathiapalan,
``Closed String Tachyons on $C/Z_N$'',
hep-th/0309029.


\bibitem{Marcus}
N. Marcus,
``Unitarity and Regularized Divergences in String Amplitudes'',
\J{\PL}{B219}{1989}{265}.



\bibitem{Dab-2}
A. Dabholkar,
``Strings on a Cone and Black Hole Entropy'',
\J{\NP}{B439}{1995}{650},
hep-th/9408098.

\bibitem{Lowe}
D. A. Lowe and A. Strominger,
``Strings Near a Rindler or Black Hole Horizon'',
\J{\PR}{D51}{1995}{1793},
hep-th/9410215.

\bibitem{Takayanagi}
T. Takayanagi and T. Uesugi,
``Orbifolds as Melvin Geometry'',
\J{JHEP}{0112}{2001}{004},
hep-th/0110099.

\bibitem{S.Weinberg}
S. Weinberg,
``Cancellation of One Loop Divergences in SO(8192) String Theory'',
\J{\PL}{B187}{1987}{278}.


\bibitem{Bardakci}
K. Bardakci and A. Konechny,
``Tachyon Condensation in Boundary String Field Theory at One Loop'',
hep-th/0105098;
``One Loop Partition Function in the Presence of Tachyon
Background and Corrections to Tachyon Condensation'',
\J{\NP}{B614}{2001}{71}.

\bibitem{Craps}
Ben Craps, Per Kraus and Finn Larsen,
``Loop Corrected Tachyon Condensation'',
\J{JHEP}{0106}{2001}{062},
hep-th/0105227.

\bibitem{Nakamura}
S. Nakamura,
``Closed-string Tachyon Condensation and the On-shell Effective Action 
of Open-string Tachyons'',
\J{\PTP}{106}{2001}{989},
hep-th/0105054.


\bibitem{Weinberg-Wu}
E. J. Weinberg and  A. Wu,
``Understanding Complex Perturbative Effective Potentials'',
\J{\PR}{D36}{1987}{2474}.

\bibitem{Callan}
C. G. Callan and S. R. Coleman,
``The Fate of the False Vacuum. II. First Quantum Corrections'',
\J{\PR}{D16}{1977}{1762}.

\end{thebibliography}
\end{document}